# Assessing Greenspace Attractiveness with ChatGPT, Claude, and Gemini: Do AI Models Reflect Human Perceptions?

Milad Malekzadeh[1*], Magdalena Biernacka[2], Elias Willberg[1], Jussi Torkko[1], Edyta Łaszkiewicz[2], Tuuli Toivonen[1]

**Abstract**

Understanding greenspace attractiveness is essential for designing livable and inclusive urban environments, yet existing assessment approaches often overlook informal or transient spaces and remain too resource-intensive to capture subjective perceptions at scale. This study examines the ability of multimodal large language models (MLLMs), ChatGPT GPT-4o, Claude 3.5 Haiku, and Gemini 2.0 Flash, to assess greenspace attractiveness similarly to humans using Google Street View imagery. We compared model outputs with responses from a geo-questionnaire of residents in Łódź, Poland, across both formal (e.g., parks, managed greenspaces) and informal (e.g., meadows, wastelands) greenspaces. Survey respondents and models indicated whether each greenspace was attractive or unattractive and provided up to three free text explanations. Analyses examined how often their attractiveness judgments aligned and compared their explanations after classifying them into shared reasoning categories. Results show high AI–human agreement for attractive formal greenspaces and unattractive informal spaces, but low alignment for attractive informal and unattractive formal greenspaces. Models consistently emphasized aesthetic and design-oriented features, underrepresenting safety, functional infrastructure, and locally embedded qualities valued by survey respondents. While these findings highlight the models' potential for scalable pre-assessment, they also underscore the need for human oversight and complementary participatory approaches. We conclude that MLLMs can support, but not replace, context-sensitive greenspace evaluation in planning practice.

---

[1] Digital Geography Lab, Department of Geosciences and Geography, University of Helsinki, Finland
[2] Social-Ecological Systems Analysis Lab, Faculty of Economics and Sociology, University of Lodz, Poland
* milad.malekzadeh@helsinki.fi

1. **Introduction**

Greenspaces provide multiple ecosystem services and benefits to human health and wellbeing, especially in urban areas where the majority of the global population lives (Frumkin et al., 2017; Markevych et al., 2017). Recent research has moved beyond questions of greenspace availability to examine their quality and attractiveness (van den Berg et al., 2015), understood here as the extent to which individuals willingly choose to visit and spend time in a greenspace that meets their needs, expectations, and preferences (Biernacka and Kronenberg, 2018). Scholars have assessed greenspace attractiveness through a range of approaches, including surveys, participatory and mental mapping and interviews (Biernacka and Kronenberg, 2018; Otto et al., 2024), as well as data-driven techniques that integrate expert evaluations with remote sensing and street-level imagery (Zhang et al., 2023).

However, evaluating attractiveness across different forms of urban greenery remains a challenge, as these spaces vary greatly in their management, appearance, and social meaning (Biernacka and Kronenberg, 2018). Formal greenspaces, such as parks, managed forests, cemeteries, and street greenery, are typically maintained by municipal authorities and designed to provide structured recreation and aesthetic value. In contrast, informal greenspaces, including vacant lots, abandoned brownfields, and self-regenerating areas, are often unmanaged, less accessible, and more diverse in their ecological and visual qualities. Although both contribute to urban resilience and ecological functioning, they tend to evoke distinct perceptions and expectations among city residents (Biernacka and Kronenberg, 2018). Understanding which features make these varied greenspaces attractive to diverse demographic groups is therefore essential for planners seeking to design livable, inclusive, and ecologically balanced cities.

Recently, generative AI tools, especially multimodal large language models (MLLM), have emerged as a promising technique to assess to assess the attractiveness and quality of greenery and more broadly of urban environments. MLLMs have shown strong capabilities in vision–language tasks, including generating evaluations and descriptions of images in response to simple textual prompts (Liang et al., 2024). To assess attractiveness, a quickly growing number of studies have compared AI evaluations to human perception and shed light on the internal reasoning processes of the MLLMs models, which largely remain obscure. For example, Malekzadeh et al., (2025) compared evaluations of urban attractiveness generated by ChatGPT-4 from street view imagery against the perceptions of residents and tourists and found a promising agreement between the assessments, but also spatial mismatches. In a similar fashion, Wedyan et al., (2025) assessed MLLMs capabilities in estimating walkability finding both alignment and notable difference between the human and computer-based assessments, while (Ki et al., 2025) proposed a framework for integrating MLLMs to support walkability assessments in planning. Building on these methodological advances in urban design and environmental quality research, it is timely to ask whether MLLMs can be used to complement people's assessments of greenspace attractiveness.

It appears clear that MLLMs have potential to address several limitations of the more established greenspace evaluation methods. First, research on greenspace attractiveness has predominantly concentrated on formal greenspaces, while informal or transient ones remain underexplored, despite their importance for everyday urban experiences. This imbalance means that public preferences for informal spaces are often overlooked in planning and policy contexts (Biernacka & Kronenberg, 2018). Second, existing assessment methods, whether based on audits (Biernacka et al., 2020) or participatory surveys and their geo-enabled extensions such as Public Participation GIS (PPGIS) (Łaszkiewicz et al., 2025), tend to be resource-intensive and difficult to scale (Knobel et al., 2019). Given these challenges, MLLMs offer a promising opportunity for scalable and relatively low-effort evaluations of greenspace attractiveness, capable of encompassing both formal and informal environments and supporting more inclusive greenspace management and planning.

Yet to date, exploring the applicability of MLLMs for evaluating human perceptions of urban environments, particularly in the context of greenspaces, remains a emerging field, largely due to their novelty. In particular, there is gap in understanding the reasoning processes of these models, specifically their ability to explain why certain environments are perceived more positively than others, rather than merely classifying them as such. In the context of greenspace attractiveness, it remains unexplored how well and reliably MLLMs can capture the multisensory nature experiences in different spatial contexts and the diverse features that affect attractiveness. Similarly, it remains open whether they can adequately represent differences in perceptions and values between demographic groups as hidden gender and other biases remain a common problem for AI models (Hall and Ellis, 2023). We also do not know whether MLLMs are more suitable for evaluating the attractiveness of large, formal greenspaces compared with more fragmented and smaller informal greenspaces.

We employed three state-of-the-art MLLMs, including ChatGPT GPT4o, Claude 3.5 Haiku, and Gemini 2.0 Flash together with a stratified geo-questionnaire collected from Lodz, Poland to compare AI-generated evaluations of urban greenspace attractiveness to human perception. More specifically, we compared respondents' binary and descriptive assessments of the attractiveness of different urban greenspace categories with the classifications and open-ended reasoning generated by three MLLMs. As the input for the AI models, we used local street view imagery collected from the locations mapped by the geo-questionnaire respondents. In our analysis, we distinguished between formal and informal greenspaces and used a role-playing approach in which the MLLMs were prompted to first employ the perspective of a generic citizen and then the socio-demographic profile of the individual participant who had evaluated the greenspace. Finally, we decoded the AI evaluations using semantic segmentation.

The main contribution of our study lies in demonstrating how multimodal large language models can approximate human reasoning about greenspace attractiveness through visual–text evaluation.

## 2. Methodology

### 2.1. Overview

While human experiences of greenspaces are inherently multisensory and shaped by direct interaction with the environment (Thibaud, 2011), AI models require a visual proxy to process and interpret such environments. To bridge this gap, we employ Street View Imagery (SVI) as a standardized visual input.

Human participants were asked to identify greenspaces in the immediate vicinity of their place of residence that they perceived as attractive or unattractive and to provide open-ended explanations for their assessments. In parallel, we designed prompts for three MLLMs—OpenAI's ChatGPT GPT-4o, Anthropic's Claude 3.5 Haiku, and Google's Gemini 2.0 Flash—using the same introductory context and framing as provided to participants. Each model was prompted in two ways: (1) from the perspective of a generic citizen of Łódź, Poland, and (2) from the perspective of a hypothetical citizen whose socio-demographic profile matched that of the participant evaluating the same greenspace. In both cases, the models generated free-text explanations for their judgments of greenspace attractiveness.

We systematically compared the explanatory reasoning of the AI models and human participants using both semantic clustering and manual categorization approaches. In addition, recognizing the proprietary and opaque nature of these commercial AI models, we conducted supplementary image analysis to explore the visual features potentially influencing the models' assessments. The following sections detail the methodological steps of this process.

### 2.2. Study Area

The study was conducted within a 500 m buffer around the Jasień and Karolewka rivers in Łódź, Poland, the country's fourth largest city, with 645,693 inhabitants in 2024 and a total area of 293 km² (Central Statistical Office of Poland, 2025). Historically, the city's rivers and surrounding forests supported the rapid development of the textile industry in the 19th century, but industrialisation also led to severe environmental degradation, with many rivers now channelled underground (Kronenberg et al., 2017). Today, green spaces in Łódź occur in a variety of forms, ranging from highly maintained parks to unmanaged wastelands (Biernacka et al., 2023). The selected river valleys capture the full spectrum of green space types and urban structures found across the city. Their western sections feature suburban housing, transport infrastructure, and extensive informal green spaces; the central sections include dense residential areas, formal parks, and industrial zones; and the eastern sections contain agricultural fields interspersed with informal greenery. Although the rivers themselves are relatively short (12.6 km and 3 km), the surrounding 30.77 km² study area, representing about 11% of the city's total area, covers an urban–peri-urban gradient characteristic of Łódź (**Supplementary Materials I - Figure S1**). Greenspaces occupy approximately 59% of this area, inhabited by around 17% of the city's population (about 112,000 residents) (Biernacka et al., 2023). This makes the study area well suited for analysing residents' perceptions of both formal and informal green spaces.

### 2.3. Human Assessment of Greenspace Attractiveness

To explore how residents perceive the attractiveness of both formal and informal UGS, we conducted a household survey among individuals living within 500 m of the Jasień and Karolewka river corridors in Łódź. The survey targeted a representative sample of residents (N = 407) and was carried out by trained interviewers from an external research company between November 2019 and January 2020.

A stratified random sampling procedure was used. The Jasień Valley study area was divided into eight strata based on territorial divisions defined by the research team, and the number of participants selected from each stratum was proportional to its estimated population size. In total, 2,000 addresses were randomly selected. Interviews were conducted face-to-face at participants' homes. If a selected resident was not available or declined participation, interviewers proceeded to the next randomly selected address within the same stratum. The survey achieved a 32.2% response rate, with an average interview duration of approximately 19 minutes.

The questionnaire included questions on the use and perception of UGS, as well as respondents' socio-economic characteristics, including education level (primary, lower secondary, vocational, secondary, higher), job status (student, employed, pensioner, unemployed, other), number of children (none, one, two or more), gender (men, women, other), and birth year. A central component was a geo-questionnaire, in which respondents identified and evaluated green spaces in the vicinity of their home, whether formal (e.g., parks, forests, allotment gardens) or informal (e.g., neighborhood green spaces, waterside green spaces, brownfields), as attractive or unattractive, and provided up to three reasons for their assessment. Each selected green space was geolocated by placing a point on an interactive map. In total, respondents identified 210 formal and 36 informal sites as attractive, and 44 formal and 72 informal sites as unattractive. The responses were originally in Polish and were translated into English using one of the Polish-speaking members of the research team.

### 2.4. AI Evaluation of Urban Greenspaces

#### 2.3.1 Data

We obtained panoramic Google Street View (GSV) imagery for the greenspaces identified by survey respondents and the areas in their immediate vicinity, using the Google Maps API authorized through an API key. For formal greenspaces, image collection was confined to the official park boundaries. In contrast, for informal greenspaces lacking clearly defined perimeters, we applied a 100-meter buffer around each point mapped by the participants to delineate the area of interest. To ensure comprehensive coverage of the surrounding environment in both cases, we added an additional 10-meter buffer to the resulting polygons. To systematically sample the spatial extent of each greenspace, we generated a regular grid of points within the buffered polygons, using a spacing of 50 m between points. This was operationalized by iteratively creating points at fixed intervals along the x and y axes within the bounding box of each polygon, and retaining only those points falling within the polygon's boundaries. Each of these interior points served as a query location for GSV image acquisition. Each GSV image was downloaded at a resolution of 640 × 640 pixels, with a horizontal field of view of 60° and a pitch angle of 0°. For each location, we assembled a panoramic 360° image by capturing six directional images oriented at 0°, 60°, 120°, 180°, 240°, and 300°, with 0° facing true north.

Given the input limitations of the AI models, a maximum of 10 images was assigned to each participant-evaluated greenspace, regardless of whether it was formal or informal. For cases, particularly in large formal parks, where more than 10 images were available, we manually selected a representative subset. This selection aimed to encompass a diverse range of park features and functions, including but not limited to social infrastructure, playgrounds, and seating areas. This approach was necessary because participants typically evaluated entire parks rather than specific locations within them, necessitating a holistic and representative visual input for the AI models.

### 2.3.2 Models and Prompts

We employed three state-of-the-art MLLMs to assess SVI: OpenAI's ChatGPT GPT-4o, Anthropic's Claude 3.5 Haiku, and Google's Gemini 2.0 Flash. These models were selected based on their widespread use, established multimodal capabilities, and the availability of both user-friendly interfaces and scalable application programming interfaces (APIs), which allowed for scalable and automated querying at the required volume.

Our prompt design aimed to replicate the human evaluation process as closely as possible. Specifically, we formulated prompts that mirrored the instructions and framing provided to human participants. However, to elicit the most contextually relevant responses from the models, we implemented a role-playing approach in the prompt design. In the first approach, each model was instructed to assume the role of a generic citizen of Łódź, which in practice meant that the prompt simply referred to a "citizen of Łódź" without providing any additional demographic details. In the second approach, the models were prompted to adopt the socio-demographic profile of the individual participant who had evaluated the same greenspace. This allowed us to investigate whether the adoption of a personalized demographic context influenced the model's evaluation and reasoning.

In both approaches, we explicitly defined the concept of "greenspace attractiveness" and instructed the models to provide (1) a binary judgment indicating whether the greenspace was attractive or unattractive, and (2) three free-text reasons explaining their evaluation. The responses were unconstrained by predefined categories to align with the open-ended nature of the human participants' answers. The exact wording of the prompts for both conditions is provided in Supplementary Material II.

All models were queried using a maximum of 10 images per location, corresponding to the models' input limitations. The temperature parameter, which influences response creativity and variability, was set to 0.5 to balance coherence and variability in the outputs. Each prompt was submitted as an independent, single-turn query; we did not employ multi-turn conversational prompting or context chaining (commonly referred to as "single-shot" prompting).

Following the retrieval of model outputs, we performed a standardized cleaning process to remove extraneous text and ensure that the responses adhered to the requested format. Some models occasionally added superfluous commentary or explanations beyond the specified structure. The final dataset consisted of six distinct responses for each location, reflecting three models and two prompt conditions per model (generic and personalized). Examples of model outputs and the extraneous explanatory text removed during the cleaning process are provided in Supplementary Materials III.

## 2.5. Comparison of Human and AI Assessments

The comparison between human and AI evaluations was conducted in two stages. First, we assessed the level of agreement between human participants and the AI models regarding whether each greenspace was perceived as attractive or unattractive on a one-to-one basis. Second, we examined the underlying reasons provided for these evaluations, comparing the explanatory statements generated by the models with those expressed by human participants.

Given that both human and AI responses were open-ended free-text, we applied two complementary analytical approaches to assess the similarity of these explanations. The first approach involved semantic clustering of the textual responses using unsupervised machine learning techniques. While the semantic clustering approach provided an initial overview of the reasoning patterns, it did not capture the full contextual meaning of the responses. This was likely due to the complexity of natural language explanations, the diversity of vocabulary used by participants and models, and the challenges inherent in clustering short, context-dependent sentences. To address this limitation and enable a more interpretable comparison, we performed a manual categorization of all explanatory reasons. The detailed methodology and results of the semantic clustering analysis are provided in Supplementary Materials IV.

### 2.4.1 Manual Categorization of Explanatory Reasons

Each reason provided by participants and AI models was reviewed and assigned to one of twelve predefined categories, which are listed in **Table 1**. These categories were developed inductively based on recurring themes observed in both the human and AI-generated responses, and refined through iterative discussion among the research team to ensure conceptual clarity and distinctiveness.

**Table 1.** Definition of reasoning categories used for manual classification of explanatory responses.

| Category | Examples of Associated Phrases |
| --- | --- |
| Access | free entry, open access, public place, fencing, private property, access to public transport, muddy ground |
| Aesthetics | cleanliness, well-kept lawns, decorative plantings, scenic view, overgrown, graffiti, industrial surroundings |
| Ambient environmental conditions | calm, peaceful, urban noise, pollution from traffic, bad smell |
| Infrastructure for play | playgrounds, swings, chess tables, lack of playgrounds, facilities in bad condition, playground looks worn |
| Infrastructure for walking | wide alleys, even paths, unpaved, crooked, narrow paths, muddy paths |
| Infrastructure for physical activity | sports fields, outdoor gyms, bike paths, training areas |
| Social infrastructure | benches, toilets, waste bins, picnic facilities, catering facilities, lack of benches |
| Safety | safe, well-lit, lack of lighting, looks unsafe, presence of dangerous people |
| Nature | trees, flowering shrubs, pond, wildflowers, dense tree cover, lack of flowers, potential for wildlife |
| Proximity | close to home, far from home |
| Landscape layout | open space, rolling hills, varied landscape, spacious area, clear focal point |

Following the categorization, we compared the distribution of these reasoning categories across human participants and the six AI model configurations (three models, each tested with two prompting conditions). The comparisons were conducted separately for greenspaces perceived as attractive and unattractive, and for formal versus informal greenspaces. This resulted in four distinct comparison groups:

- Attractive – Formal greenspaces
- Attractive – Informal greenspaces
- Unattractive – Formal greenspaces

- Unattractive – Informal greenspaces

In addition to comparing the frequency distributions across these groups, we also generated co-occurrence matrices for the AI model outputs. These matrices illustrate how often different reasoning categories appeared together within individual model responses, helping to reveal patterns of combined reasoning that the models applied when evaluating greenspace attractiveness. This approach provided further insight into the complexity of the AI-generated explanations and allowed for a nuanced comparison against human reasoning patterns.

### 2.6. Decoding AI Models' Outputs

As the AI models used in this study are proprietary systems, their internal architectures and multimodal reasoning processes, particularly in the context of visual perception and scene evaluation, remain opaque. While these models demonstrate considerable capabilities in object recognition and spatial reasoning (Chen et al., 2024), the specific mechanisms through which they derive judgments of greenspace attractiveness are not transparent. To investigate this black-box behaviour, we conducted an exploratory image analysis focusing on the visual features present in the Street View imagery used for AI evaluation.

We first applied semantic segmentation to quantify the relative proportions of key environmental features within each image. Specifically, we extracted the pixel coverage of five primary object classes: greenery, sky, road, sidewalk, and buildings. For each class, we calculated its proportion relative to the total pixel count of the image, yielding a normalized measure of its visual prominence within the [0, 1] range. While semantic segmentation performs well in identifying larger environmental elements, it is less appropriate for smaller, countable objects, such as benches or individual people, where frequency, rather than area, is a more meaningful metric. To capture these smaller-scale but perceptually relevant elements, we additionally employed instance segmentation targeting five object categories: vehicles, decorative elements, urban furniture, people, and lighting/signage. This allowed us to count the number of distinct objects detected within each image. To ensure comparability between the count-based and area-based measures, we normalized all values using a min–max scaler, rescaling object counts to the [0, 1] range consistent with semantic segmentation proportions. Because min–max scaling is sensitive to outliers, values exceeding the mean + 2 standard deviations were excluded prior to normalization. A full mapping of ADE20K classes to the aggregated categories used in both semantic and instance segmentation and examples of segmented scenes is provided in Supplementary Material V.

The semantic segmentation task was completed using the NVIDIA SegFormer-B5 model, specifically the "nvidia/segformer-b5-finetuned-ade-640-640" variant (Xie et al., 2021). For the instance segmentation, we employed the "oneformer_ade20k_dinat_large" variant of the OneFormer model (Jain et al., 2023). Both models are pre-trained on the ADE20K dataset (Zhou et al., 2017) and are publicly available.

The segmentation analysis was conducted on the Finnish CSC's Puhti supercomputer, using Nvidia V100 GPUs. The resulting pixel-level classifications for each image were aggregated and used as explanatory features to explore associations between visual content and the reasoning categories generated by the AI models.

## 3. Results

### 3.1. Binary-Level Agreement Between AI Models and Participants

Agreement between AI models and participants in classifying greenspaces as attractive or unattractive varies considerably across greenspace types and attractiveness categories (**Table 2**). Agreement is consistently high for attractive formal greenspaces, with Claude (92.4%) showing the strongest alignment. This suggests that models reliably recognize attractive features in structured parks and formal greenspaces.

**Table 2.** Percentage agreement between each model and participants by greenspace type

| Model | Attractive – Formal | Attractive – Informal | Unattractive – Formal | Unattractive - Informal |
|---|---|---|---|---|
| GPT-4o (Generic) | 90.0 | 19.4 | 20.5 | 83.3 |
| GPT-4o (Personalized) | 89.0 | 19.4 | 25.0 | 83.3 |
| Gemini (Generic) | 89.5 | 19.4 | 25.0 | 84.7 |
| Gemini (Personalized) | 84.8 | 16.7 | 25.0 | 87.5 |
| Claude (Generic) | 91.5 | 36.1 | 14.3 | 73.9 |
| Claude (Personalized) | 92.4 | 33.3 | 13.6 | 72.2 |

In contrast, agreement drops sharply for attractive informal greenspaces, ranging from 16.7% to 36.1%, with Claude performing slightly better than the other models. This discrepancy likely reflects two factors: first, informal greenspaces often lack the conventional design elements that models are trained to associate with attractiveness; second, SVI often captures these spaces from street angles where the actual greenspace may be obscured or visually disconnected from what participants experienced.

For unattractive greenspaces, the pattern reverses. Agreement is substantially higher for unattractive informal greenspaces (up to 87.5%), suggesting that models effectively detect negative cues when these are visible from the street. However, for unattractive formal greenspaces, agreement remains low (13.6%–25.0%), indicating that models tend to default to a positive assessment of parks, potentially overlooking contextual issues less apparent in imagery.

Differences between generic and personalized prompts are marginal across all models and categories (see **Supplementary Materials VI – Table S5**), suggesting that adding socio-demographic context to the prompts does not meaningfully influence the models' evaluations. Taken together, these findings indicate that MLLMs perform well when evaluating visually clear, formal environments and when detecting unattractiveness in informal greenspaces. However, they struggle with the subjective and often less visually explicit attractiveness of informal greenspaces, partly due to the limitations of street-level imagery as the input source.

### 3.2. Manual Categorization of free-text responses

When examining the distribution of categories across the models, several patterns emerge (**Figure 1**). Comparing AI-generated and participant responses reveals several key discrepancies. The most notable is safety, which participants frequently mention, particularly for unattractive greenspaces, but which AI models almost entirely overlook. For attractive formal greenspaces, AI models emphasize infrastructure for walking and nature, whereas participants place comparatively less emphasis on these features. In attractive informal greenspaces, AI models focus on aesthetics, while participants more often mention infrastructure for physical activity and play, highlighting the importance of functional use over visual appeal.

In unattractive greenspaces, aesthetics emerges as the dominant reason across both formal and informal areas for all groups. However, while safety remains a concern for participants, particularly in formal greenspaces, it is largely absent from the models' explanations. In contrast, AI models frequently cite other factors, such as infrastructure for walking, nature, and social infrastructure, that participants mention far less in unattractive contexts.

AI models generally display similar reasoning patterns across both attractive and unattractive formal greenspaces. A comparable structure is found in informal greenspaces, although Gemini stands out for attributing more importance to aesthetics and paying less attention to social infrastructure compared to GPT-4o and Claude. Differences between generic and personalized prompting remain minimal across all models.

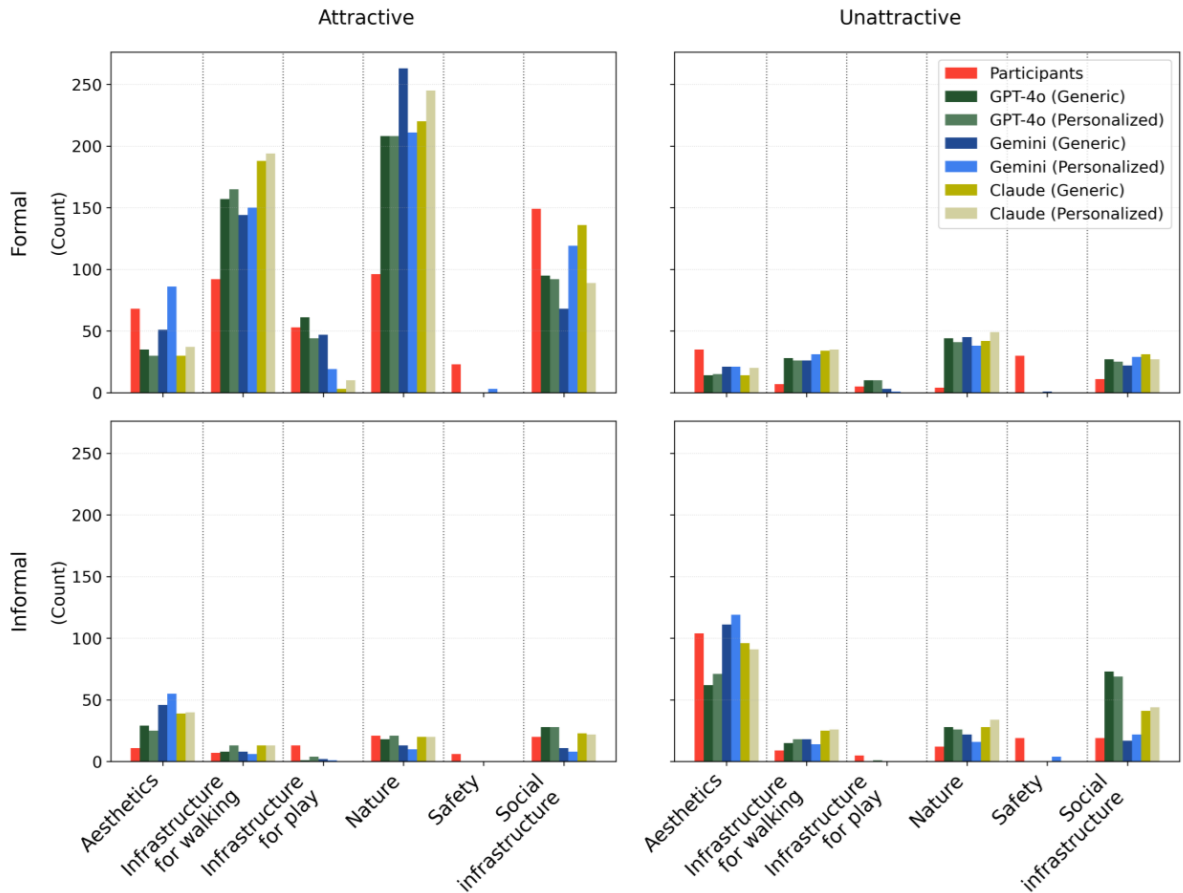

**Figure 1.** Distribution of selected reasoning categories across participants and AI models for attractive and unattractive greenspaces, separated by formal and informal spaces. Full results for all reasoning categories is provided in Supplementary Materials VII.

After analysing the overall distribution of reasoning categories, we further examined how often AI models selected the same reasoning category as participants for each point (**Figure 2**). While agreement levels are generally low across most categories, a few key areas show stronger alignment with human reasoning.

GPT-4o shows its strongest alignment in formal attractive greenspaces, particularly in the categories of Infrastructure for Walking (73–78%) and Nature (79–81%). This suggests that GPT-4o reliably detects and values conventional design elements typically associated with formal park attractiveness. In unattractive informal greenspaces, GPT-4o also shows high agreement on Aesthetics (72–78%), reflecting its focus on visible environmental degradation. Social Infrastructure is another category where GPT-4o shows high agreement in informal greenspaces (up to 100%), although this alignment is less pronounced in formal areas. However, GPT-4o rarely matches human reasoning in categories such as Safety, Infrastructure for Physical Activity and Play, and Access, where agreement remains consistently low.

Gemini performs similarly to GPT-4o in formal attractive greenspaces, with high alignment for Nature (80–86%) and moderate performance for Infrastructure for Walking (63–67%). In unattractive informal greenspaces, Gemini achieves even higher agreement in Aesthetics (83–87%). In contrast, agreement in Social Infrastructure remains lower and more variable across conditions.

Claude achieves high agreement levels overall, particularly in Infrastructure for Walking (89–92%) and Nature (85–87%) in attractive formal greenspaces. In unattractive informal spaces, Claude also performs well, showing high agreement for Aesthetics (75–78%) and notably strong alignment in Social Infrastructure, reaching up to 82% agreement in the personalized setting. These results suggest that Claude is relatively effective at capturing both positive environmental qualities in formal greenspaces and negative cues, such as lack of amenities, in informal environments.

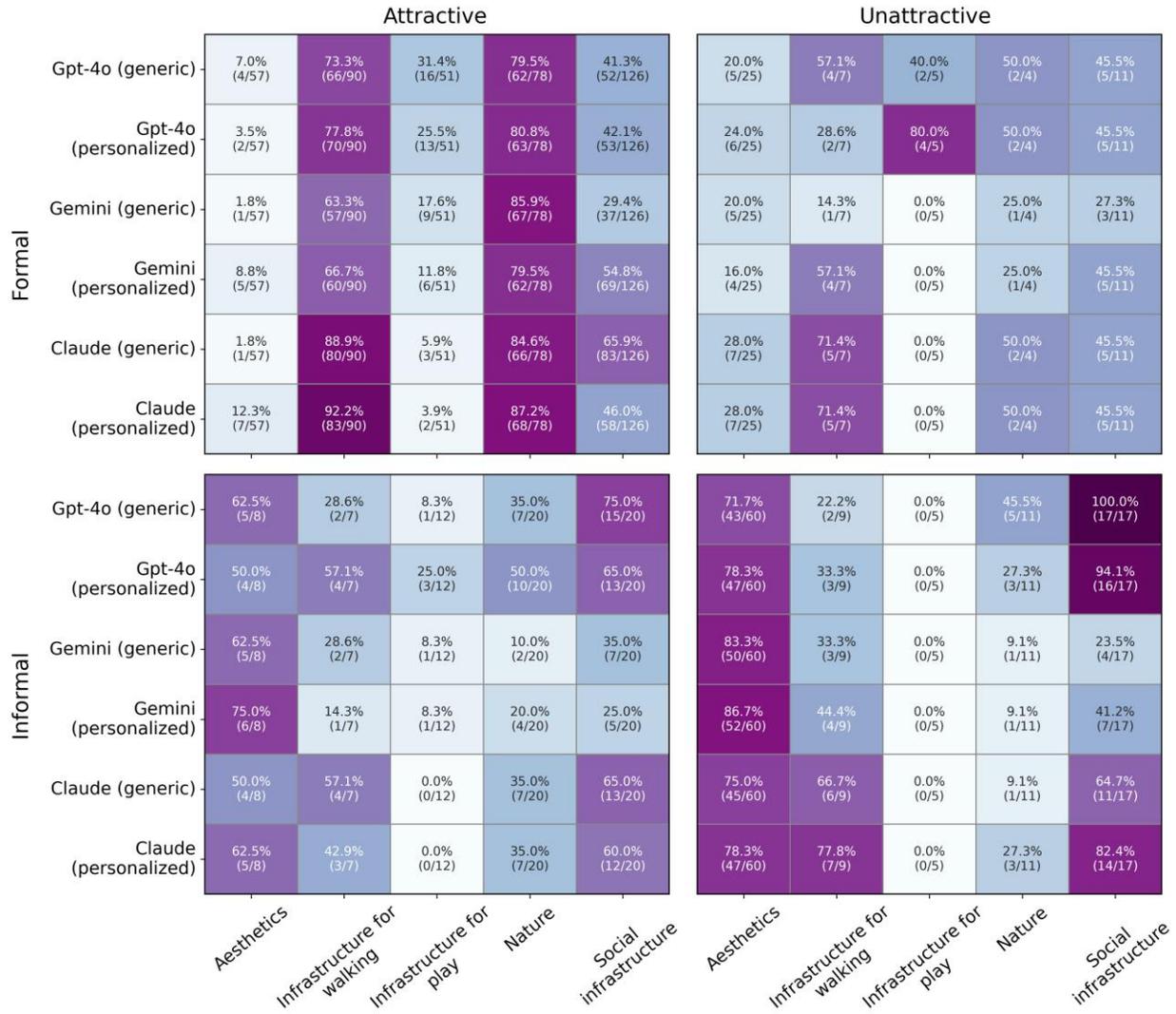

**Figure 2**. Agreement between AI models and participants on selected reasoning categories across greenspace types. Each cell shows the percentage of cases where an AI model selected the same reasoning category as participants for a given greenspace, reported separately for attractive and unattractive, formal and informal spaces. In brackets, the absolute counts indicate how many times the model matched the participant's reasoning over the total number of times participants mentioned that category. Darker shades indicate higher levels of agreement. Columns are shown in gray where participants did not mention the category for that environment. Full results for all reasoning categories is provided in Supplementary Materials IX.

To explore how reasoning categories co-occur within the AI-generated explanations, we analyzed the normalized co-occurrence matrices across models and prompt settings (**Supplementary Materials X - Figure S7**). The results reveal consistent patterns across both generic and personalized prompts, with only minor variations between models (see Supplementary Materials X).

### 3.3. Association Between Image Features and AI Models Outputs

When examining the image features associated with each reasoning category, we observe a high level of consistency across all models (**Figure 3**). Here, we present results for four selected categories; the full set of plots is available in Supplementary Materials XI. Some variations across Ambient Environmental Conditions, Infrastructure for Physical Activity, and Safety categories likely reflect the smaller sample sizes for these categories.

For most categories, Greenery is a dominant visual element, particularly in Nature-related explanations, but it surprisingly plays a smaller role in Aesthetics and Access. Buildings appear more frequently in categories such as Aesthetics, Access, and Landscape Layout but remain low in others. Roads show moderate levels across categories but are highest in access-related reasoning. Sidewalks are generally less prevalent, except for moderate levels in Access and Infrastructure for Physical Activity. Lighting & Signs are relatively low overall but more noticeable in Access reasonings. People consistently appear at moderate levels across all categories. Furniture peaks in Infrastructure for Play but is minimal in Aesthetics and Access. Finally, Vehicles are most common in access-related reasoning and least visible in others.

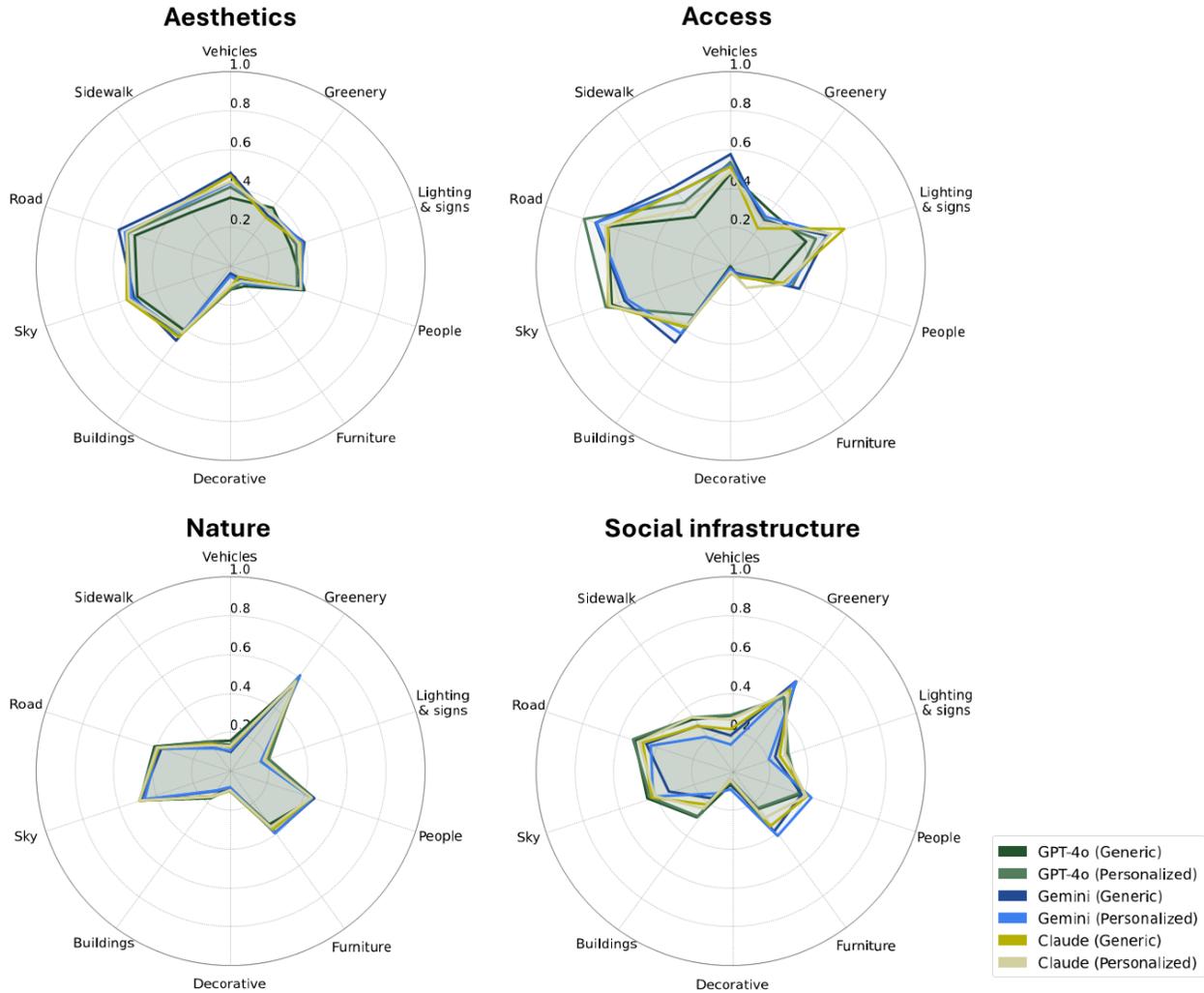

**Figure 3.** Distribution of image features across selected reasoning categories. Full results for all reasoning categories are available in Supplementary Materials XI. Each radar chart shows the normalized average proportion of visual features detected in the images linked to each reasoning category, as interpreted by the AI models.

## 4. Discussion

Our study compared greenspace attractiveness evaluations generated by three state-of-the-art multimodal large language models with human perceptions collected through a local geo-questionnaire in Łódź, Poland. Based on SVI and open-ended evaluation tasks, we assessed both agreement on binary attractiveness judgments and similarity in the explanatory reasoning across formal and informal greenspaces, revealing systematic biases in how AI interprets formal and informal urban nature.

We found that AI–human agreement followed a clear pattern across greenspace types. In formal greenspaces, models consistently tended to classify areas as attractive, as they reliably identified conventional park features such as greenery, walking infrastructure, and scenic layout. This tendency resulted in high agreement for attractive formal greenspaces but low agreement for formal greenspaces that participants rated as unattractive. In informal greenspaces, where attractiveness often depended on functional qualities or experiential values less visible in imagery, models typically judged these areas as unattractive. Consequently, agreement was low for attractive informal sites and high for unattractive ones. Across contexts, models overemphasized aesthetic cues and underrepresented functional infrastructure, especially in informal spaces, reflecting a potential bias towards curated, design-oriented environments. Furthermore, the negligible differences between generic and personalized prompts suggest that model outputs remain primarily image-driven, with limited sensitivity to socio-demographic framing. Beyond these socio-demographic factors, human perceptions of green spaces are also shaped by individual experiences, abilities, and emotional associations, which influence what people notice and value (Wolff et al., 2022). This inherent subjectivity remains difficult for AI models to capture.

The strong agreement we observed for attractive formal greenspaces aligns with Malekzadeh et al., (2025), who found that MLLMs achieve their highest performance when visual inputs conform to conventional design features. This pattern suggests that such models may be implicitly tuned, through their training data, to recognize and value curated, well-maintained environments that dominate online image repositories, where informal spaces are rarely depicted or explicitly labelled as greenspaces (Pietrzyk-Kaszyńska et al., 2017). The overemphasis on aesthetics in our results reflects patterns reported in recent computer vision–based urban form studies (Vanky and Le, 2023), where visual appeal frequently outweighs functional or social qualities in model outputs. The low agreement observed for attractive informal greenspaces supports the critiques of Biernacka and Kronenberg, (2018) and van den Berg et al., (2015), who argue that the willingness to spend time in a greenspace, and its perceived attractiveness, often stems from experiential, functional, and locally embedded qualities, as well as prior knowledge or perceptions of the place, that are not readily discernible in visual imagery alone.

Unlike findings from text-only LLM research where user profiling can influence responses (Marvin et al., 2024), our study found minimal effect of socio-demographic framing in visual–text evaluations, suggesting that current MLLMs rely predominantly on visual pattern recognition with limited integration of social context. The concentration of AI reasoning within a narrow set of categories such as nature, aesthetics, infrastructure for walking, and social infrastructure, supports critiques by (Belaroussi, 2025; Malekzadeh et al., 2025) that MLLMs employ a constrained conceptual vocabulary when interpreting urban spaces. While such consistency may facilitate standardization, it risks overlooking place-specific cultural or historical factors shaping perception. Without targeted training and prompt design, these models may perpetuate a "design-centric" bias rooted in their reliance on visual cues alone. Lacking awareness of social or historical context, they equate what is visually orderly or well-maintained with positive quality, while failing to recognize less visible dimensions of urban equity, such as safety, accessibility, inclusion, or cultural significance, that shape how places are actually experienced.

Despite their current limitations, our findings point to the growing potential of multimodal AI models as supportive tools in urban and landscape planning. While they cannot capture the lived, multisensory, or social dimensions of place, such models can already assist planners in identifying visual patterns of neglect or appeal, mapping perceived attractiveness at scale, and simulating how design modifications might influence public perceptions. Off-the-shelf AI systems can lower the technical threshold for integrating data-driven insights into planning, supporting more iterative, participatory, and visually grounded decision-making (Fu et al., 2025; Malekzadeh, 2026). Working alongside local expertise and citizen input, these models can support planners by assessing practical design options and creating clear visual or textual previews, helping planners align spatial design with the diverse ways people experience and value urban greenery.

## 5. Limitations and Future Research

Our use of SVI restricts evaluations to street- or route-visible perspectives, which may obscure or misrepresent key qualities of informal or enclosed greenspaces. A more comprehensive assessment would benefit from incorporating more inclusive visual coverage. Another limitation relates to the image cap imposed by the AI models, which allowed us to input only ten images per greenspace regardless of its size. For larger greenspaces, different respondents may have evaluated the same site differently because they experienced distinct sections of it; however, as these were officially classified under a single park name, participants referred to them collectively. Although for these sites, we manually selected images to capture the most salient and diverse features, many areas inevitably remained undocumented, potentially affecting the results. Future MLLMs with higher input capacity could accommodate more extensive imagery, providing a fuller representation of large and spatially heterogeneous greenspaces. Finally, our analysis was limited to a single city, which constrains the generalizability of the findings. Replicating the study across cities and countries with different types of green infrastructure, cultural expectations, and perceptual norms would help assess whether the observed AI–human agreement patterns hold in other contexts.

## 6. Conclusion

Our comparison of AI-generated and human evaluations of greenspace attractiveness shows both the promise and the present constraints of multimodal large language models in capturing the qualities that make urban nature inviting. The models proved capable of aligning with human perceptions when visual cues were clear and conventional, yet struggled when attractiveness depended on more subtle, functional, or experiential aspects. This gap is expected; human judgments are shaped by multisensory, cultural, and contextual knowledge that remains largely inaccessible to current AI systems relying solely on imagery.

The key takeaway is that MLLMs can already serve as a practical, scalable pre-assessment tool, particularly for identifying strengths and weaknesses in visually explicit environments. Yet, for decisions that demand a fuller understanding of how people experience urban nature, they should be embedded within participatory and context-rich approaches. As these technologies improve, their greatest value for planning and landscape management will come from complementing, rather than replacing, the diverse human perspectives that ultimately define what makes a greenspace truly attractive.

**Data availability**

Due to confidentiality agreements, individual level participant data are not publicly available. The imagery used in this study is proprietary Google Street View data and, in accordance with Google's terms of use, cannot be shared or redistributed.

**Declaration of conflicting interest**

The author(s) declared no potential conflicts of interest with respect to the research, authorship, and/or publication of this article

**Ethics approval statement**

The survey was conducted in accordance with the relevant guidelines of the Rector's Commission for Research Bioethics at the University of Łódź and the Personal Data Protection Act of 10 May 2018. In compliance with these guidelines, the study did not require additional approval from the Bioethics Committee, as it did not involve sensitive personal data, medical interventions, or vulnerable populations. All participants were fully informed about the purpose and scope of the research, their rights as participants, and the measures taken to ensure data confidentiality. Informed consent was obtained from all respondents prior to participation.

**Funding statement**

This work was supported by the National Science Centre, Poland; National Science Centre [grant number 2024/08/X/HS4/00783]. This study is also part of the GREENTRAVEL project (2023-2027) funded by the European Union (ERC, project 101044906). Views and opinions expressed are however those of the authors only and do not necessarily reflect those of the European Union or the European Research Council Executive Agency. Neither the European Union nor the granting authority can be held responsible for them.

Supplementary Material for the article:

**TITLE**

Assessing Greenspace Attractiveness with ChatGPT, Claude, and Gemini: Do AI Models Reflect

Human Perceptions?

**Authors:**

Milad Malekzadeh, Magdalena Biernacka, Elias Willberg, Jussi Torkko, Edyta Łaszkiewicz, Tuuli Toivonen

# Contents

**Supplementary Materials I – Study Area**

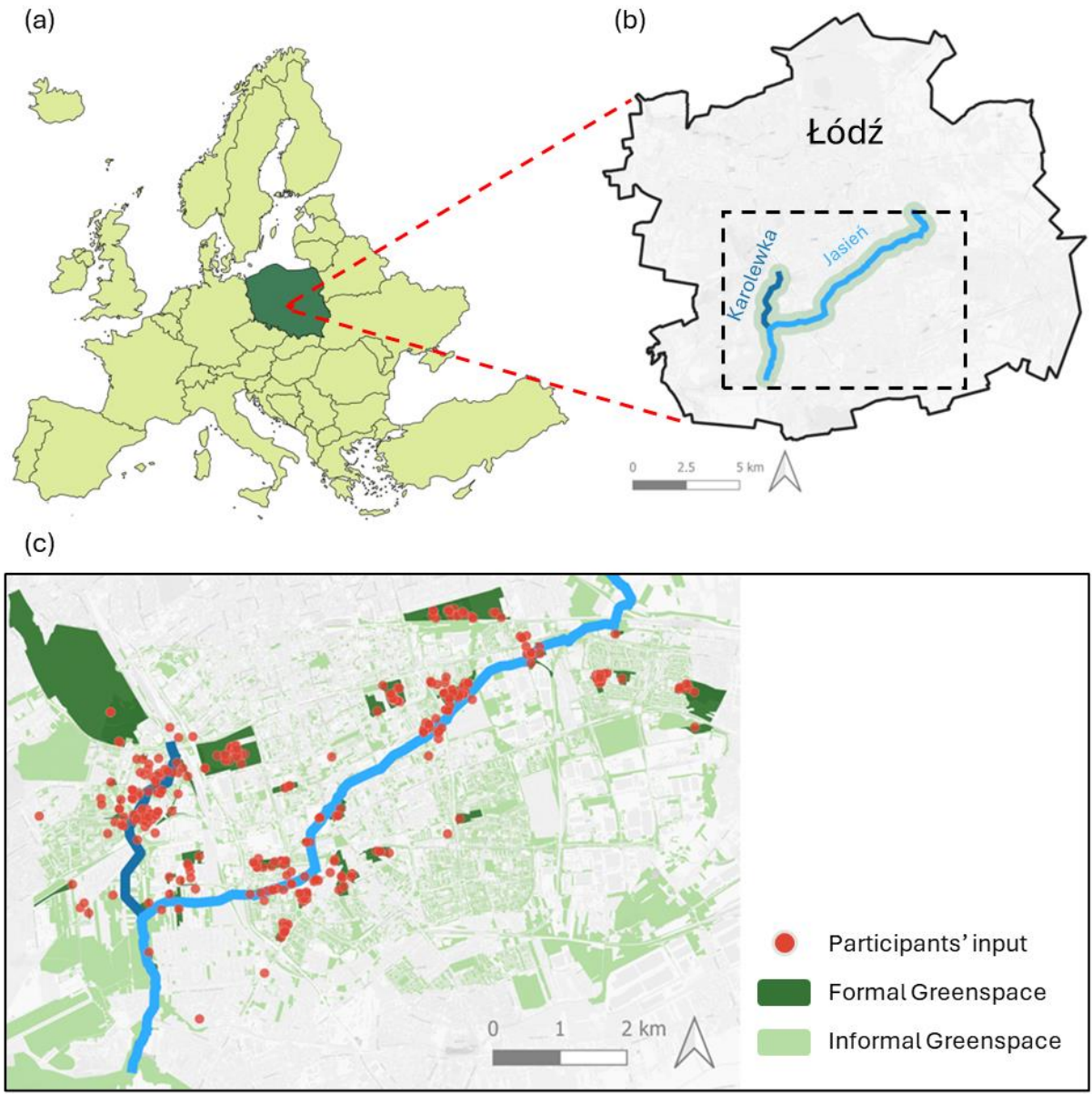

**Figure S1.** Study area covering the Jasień and Karolewka river valleys in Łódź, Poland. (a) Location of Poland within Europe; (b) location of the study area within Łódź; (c) distribution of greenspaces and participant inputs. Formal greenspaces are shown in dark green, informal

greenspaces in light green, and red dots indicate locations of greenspaces evaluated by participants.

**Supplementary Materials II – AI Models Prompts**

*Generic Prompt*

Consider yourself a citizen of the city of Lodz. You will be provided with a set of 10 panoramic images representing a single green space located within an urban area. This task involves evaluating the attractiveness of the green area depicted in these images. It is important to understand that this evaluation compares only green spaces to other green spaces; no urban cores, buildings, or non-green areas are included.

This evaluation is part of a broader survey aimed at examining the preferences of the residents of Lodz regarding the attractiveness of nearby green spaces. By green areas, we mean a variety of spaces, including lawns in front of apartment buildings, private gardens, parks, wastelands, large forests, and vast fields and meadows. We aim to understand whether people use green spaces, and if so, which ones, and to explore factors influencing their perceived attractiveness.

For clarity, attractiveness here means that you would be eager to use the area and spend time there.

Please review all 10 panorama images carefully, then respond strictly in the following structured format:

['attractive' or 'unattractive', positive: ['reason 1', 'reason 2', 'reason 3'], negative: ['reason 1', 'reason 2', 'reason 3']]

- First, explicitly state whether you find the green area attractive or unattractive overall.
- Next, list exactly three positive reasons contributing to its attractiveness.
- Finally, list exactly three negative reasons detracting from its attractiveness.

Each listed reason should be brief and concise (maximum 10 words each).

Example:

['attractive', positive: ['lush greenery', 'well-maintained paths', 'open views'], negative: ['limited seating areas', 'poor lighting', 'no water features']]

Even if the images are unclear, incomplete, repetitive, or ambiguous, you must still complete the task by making a reasonable judgment based on what you can see. Do not state that you are unable to make a judgment. Always follow the response format exactly as requested.

*Personalized Prompt*

Consider yourself a citizen of the city of Lodz with the following characteristics:

- Gender: ____
- Age: ____
- Education level: ____ (Primary education, Lower secondary, Vocational, Secondary, Higher education)
- Employment status: ____ (Student, Employed, Pensioner, Unemployed, Other status)
- Having children: ____ (No children, One child, Two or more children)

You will be provided with a set of 10 panoramic images representing a single green space located within an urban area. This task involves evaluating the attractiveness of the green area depicted in these images. It is important to understand that this evaluation compares only green spaces to other green spaces; no urban cores, buildings, or non-green areas are included.

This evaluation is part of a broader survey aimed at examining the preferences of the residents of Lodz regarding the attractiveness of nearby green spaces. By green areas, we mean a variety of spaces, including lawns in front of apartment buildings, private gardens, parks, wastelands, large forests, and vast fields and meadows. We aim to understand whether people use green spaces, and if so, which ones, and to explore factors influencing their perceived attractiveness.

For clarity, attractiveness here means that you would be eager to use the area and spend time there.

Please review all 10 panorama images carefully, then respond strictly in the following structured format:

['attractive' or 'unattractive', positive: ['reason 1', 'reason 2', 'reason 3'], negative: ['reason 1', 'reason 2', 'reason 3']]

- First, explicitly state whether you find the green area attractive or unattractive overall.
- Next, list exactly three positive reasons contributing to its attractiveness.
- Finally, list exactly three negative reasons detracting from its attractiveness.

Each listed reason should be brief and concise (maximum 10 words each).

Example:

['attractive', positive: ['lush greenery', 'well-maintained paths', 'open views'], negative: ['limited seating areas', 'poor lighting', 'no water features']]

Even if the images are unclear, incomplete, repetitive, or ambiguous, you must still complete the task by making a reasonable judgment based on what you can see. Do not state that you are unable to make a judgment. Always follow the response format exactly as requested.

**Supplementary Materials III - Examples of AI-generated responses before and after text cleaning**

The examples below illustrate some of the model outputs prior to cleaning. In these cases, each response began with a structured component (retained for analysis) followed by a narrative paragraph that was removed during preprocessing.

**Table S1.** Examples of AI-generated responses that had additional explanatory text.

| Example | Structured response (retained) | Additional explanatory text (removed) |
|---|---|---|
| 1 | ['attractive', positive: ['diverse tree landscape', 'well-maintained walking paths', 'multiple recreational areas'], negative: ['some dry grass patches', 'limited seating options', 'inconsistent ground cover']] | The green space appears to be a park-like area with several attractive features. The landscape is rich with mature trees providing shade and visual interest. The walking paths are clear and well-maintained, allowing easy navigation through the space. There are also multiple recreational areas, including playgrounds and benches, which enhance the area's usability. However, the space is not perfect. Some areas show dry grass patches, which might reduce the visual appeal. The number of seating options seems limited, and the ground cover varies in quality, which could detract from the overall aesthetic and comfort of the space. |
| 2 | ['attractive', positive: ['well-maintained walking paths', 'diverse recreational areas', 'green landscape with mature trees'], negative: ['proximity to urban infrastructure', 'limited seating options', 'some graffiti on walls']] | The green space appears to be a thoughtfully designed urban park in Łódź with several positive attributes. The walking paths are clearly defined and well-constructed, providing easy navigation through the area. There are multiple recreational zones, including a playground and what seems to be a skate park, offering activities for different age groups. The landscape is green and features mature trees that provide natural beauty and shade. However, the space is closely surrounded by apartment buildings and urban infrastructure, which might reduce the sense of being in a natural environment. There are relatively few benches or seating areas visible, which could limit relaxation opportunities. Additionally, there are some graffitied walls, which might detract from the overall aesthetic appeal of the space. |
| 3 | ['unattractive', positive: ['some green areas with grass', 'trees provide some natural element', 'yellow wildflowers | The green space appears to be primarily a utilitarian area surrounding administrative or residential buildings, with minimal design consideration for human enjoyment. |

present'], negative: ['dominated by parking spaces', 'concrete walls create harsh environment', 'lack of recreational spaces']] | While there are some green elements like grass, trees, and wildflowers, the space is overwhelmingly characterized by parking lots, concrete walls, and a generally uninviting urban infrastructure. The area lacks clear recreational potential and seems more functional than welcoming.

**Supplementary Materials IV - Semantic Clustering of Explanatory Reasons**

To compare the explanatory reasons provided by participants and AI models, we employed a semantic clustering approach. This involved transforming the free-text responses into vector representations, applying dimensionality reduction for visualization and clustering, and subsequently grouping similar explanations into thematic clusters.

We first embedded the textual responses using the Sentence-BERT model, specifically the "all-mpnet-base-v2" variant (Reimers and Gurevych, 2019). This transformer model was selected after preliminary tests comparing multilingual and English-only models, including LaBSE (Feng et al., 2020) and LASER (Artetxe and Schwenk, 2019), which showed limited performance on our Polish-English mixed corpus due to reduced accuracy on smaller or domain-specific datasets. The all-mpnet-base-v2 model provided a better semantic representation for our predominantly English AI outputs and manually translated participant responses.

Following vectorization, we explored several dimensionality reduction techniques for preparing the data for clustering and visualization, including UMAP (McInnes et al., 2018) and t-SNE (Maaten and Hinton, 2008). Based on our preliminary analysis, we selected t-SNE due to its superior separation of semantically distinct clusters. The final t-SNE parameters were set to *perplexity = 20*, *learning_rate = 100*, and *initialization = PCA*, optimized for the structure of our dataset.

For clustering, we experimented with various algorithms and determined that K-Means clustering with a pre-specified number of 10 clusters yielded the most interpretable and non-overlapping groupings of reasons. The choice of 10 clusters was based on visual inspection of the t-SNE plots and preliminary runs that revealed this number captured the key thematic patterns present in both human and AI-generated explanations. Each cluster was then qualitatively examined and labelled based on the predominant reasoning patterns observed in the grouped responses (**Table S2)**. These categories encompass key themes such as natural features, social infrastructure, aesthetics, and others. For clarity, abbreviated labels are used in the visualizations. Some clusters combine multiple themes, reflecting the limitations of sentence-level embeddings that capture word-level similarity but may overlook contextual nuance. For example, the cluster "Sa, Ae" merges safety- and aesthetics-related explanations. Similarly, broader categories like "Nature" include more specific subgroups, such as references to shade or biodiversity.

**Table S2.** Reasoning categories derived from semantic clustering and used for labeling clustered explanations.

| Name of the cluster | Abbrevation | Definition |
| --- | --- | --- |
| **Nature** | N | The cluster includes all natural elements. |
| + Shade | -I | The cluster includes all elements associated with leaves and trees that create natural shade |
| + Biodiversity | -II | The cluster includes all natural elements related to the presence of trees, plants, flowers, and description of trees |
| + Shade and biodiversity | -III | The cluster includes all natural elements combined with the provision of shade |
| + Biodiversity and landscape | -IV | The cluster includes all natural elements and landscape features, e.g., varied landscapes |
| + Shade and aesthetics | -V | The cluster includes all natural elements providing shade and water elements providing an interesting and calming view |
| + Aesthetics | -VI | The cluster includes descriptions related to overgrown space, lack of grass, flowers, greenery |
| **Social Infrastructure** | So | The cluster includes all social infrastructure elements. |
| + Seating | -I | The cluster includes social infrastructure for seating, e.g., whether there are benches, rest areas |
| + Walking | -II | The cluster includes social infrastructure for walking, e.g., whether there are paths |
| + Maintanance | -III | The cluster includes descriptions related to the maintenance of social infrastructure, e.g., well-maintained paths |
| + For children | -IV | The cluster includes elements like playgrounds, play area |
| + Seating and public | -V | The cluster includes elements like benches and descriptions related to space accessibility |
| + Walking and landscape | -VI | The cluster includes elements like walking paths and descriptions related to landscape, e.g., diversity of terrain |
| + Play and sport | -VII | The cluster includes elements like playgrounds, outdoor gym, skatepark, table tennis, playing field |
| **Aesthetics** | AE | The cluster includes descriptions related to the general aesthetics |
| + Gray infrastructure | -I | The cluster includes descriptions related to the domination of buildings, roads, cars, parking lots, concrete garages |
| Landscape | L | The cluster includes all landscape features, e.g., varied landscapes |
| Atmosphere | At | The cluster includes all elements connected with atmosphere, e.g., animals that may cause concern, e.g., rats, wild birds; former cemetery site; peaceful atmosphere |
| Proximity | P | The cluster includes all elements connected with location, e.g., proximity to home, church; location in the city |
| Recreation | R | The cluster includes descriptions related to recreation, e.g., multiple recreational areas, potential for recreation; lack of recreational facilities |
| Access | Ac | The cluster includes descriptions related to all elements that may affect ease of access, e.g., proximity to roads, proximity to buildings, proximity to railway tracks |
| Safety | Sa | The cluster includes descriptions related to safety, e.g., it is safe; it is unsafe due to lack of lighting, it is unsafe due to people drinking alcohol |

When analyzing the participant responses (**Figure S2**), we find that attractive greenspaces are predominantly associated with social infrastructure and natural features, sometimes combined with aesthetics, safety, or proximity to everyday life. In contrast, unattractive greenspaces are more frequently explained by negative aesthetic qualities, such as poor maintenance or dominance of grey infrastructure, sometimes combined with safety concerns, lack of social infrastructure, or degraded natural elements.

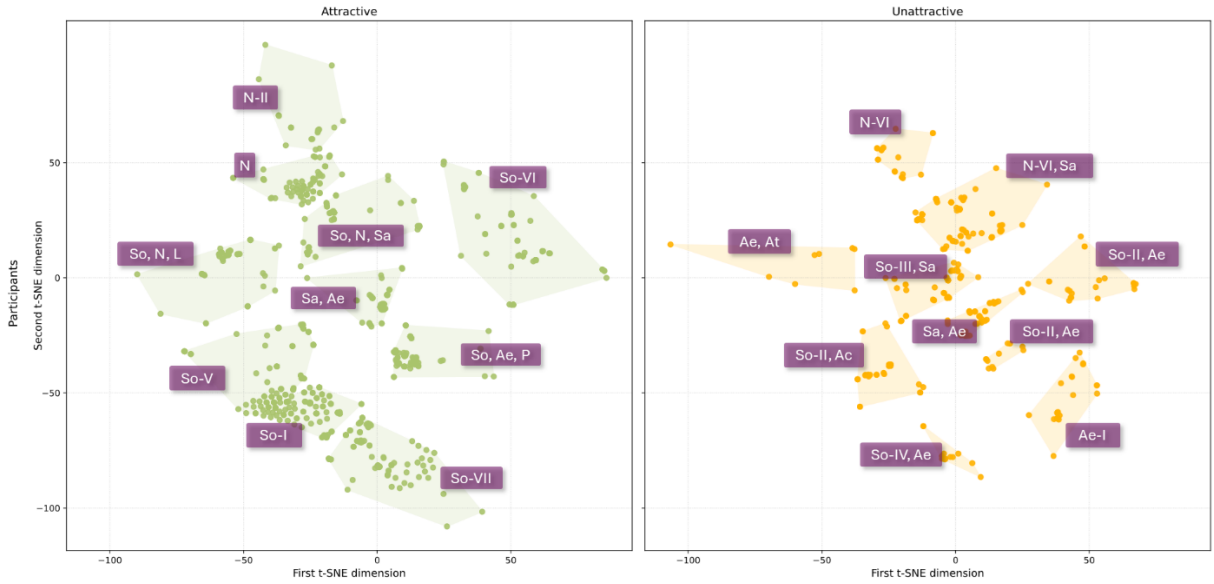

**Figure S2**. Semantic clustering of participants' explanations for attractive and unattractive greenspaces. t-SNE plots display the distribution of clustered free-text reasons provided by participants. Each cluster is labeled with its dominant themes according to **Table S2**.

When comparing participants cluster patterns to the AI-generated clusters (**Figure S3**), notable differences emerge. The t-SNE plots show that AI models and participants occupy different areas in semantic space, indicating that the wording and framing of explanations differ between them. Despite these differences, the reasoning categories used by the models follow a broadly similar structure. For attractive greenspaces, all models, GPT-4o, Gemini, and Claude, tend to emphasize Nature and Social Infrastructure, mirroring the human evaluations. However, for unattractive greenspaces, there is a clear shift in model reasoning from Nature to Aesthetics, suggesting that models focus more heavily on visible deficiencies like poor maintenance or urban clutter when making negative evaluations. Compared to participants, models also place less emphasis on social infrastructure and safety concerns in unattractive greenspaces, potentially reflecting their reliance on visual features alone without the experiential context humans bring. Additionally, unattractive greenspaces in the AI responses show greater references to Access and Recreation limitations, particularly in both generic and personalized prompts. This suggests that models often explain unattractiveness by highlighting missing recreational opportunities or poor connectivity.

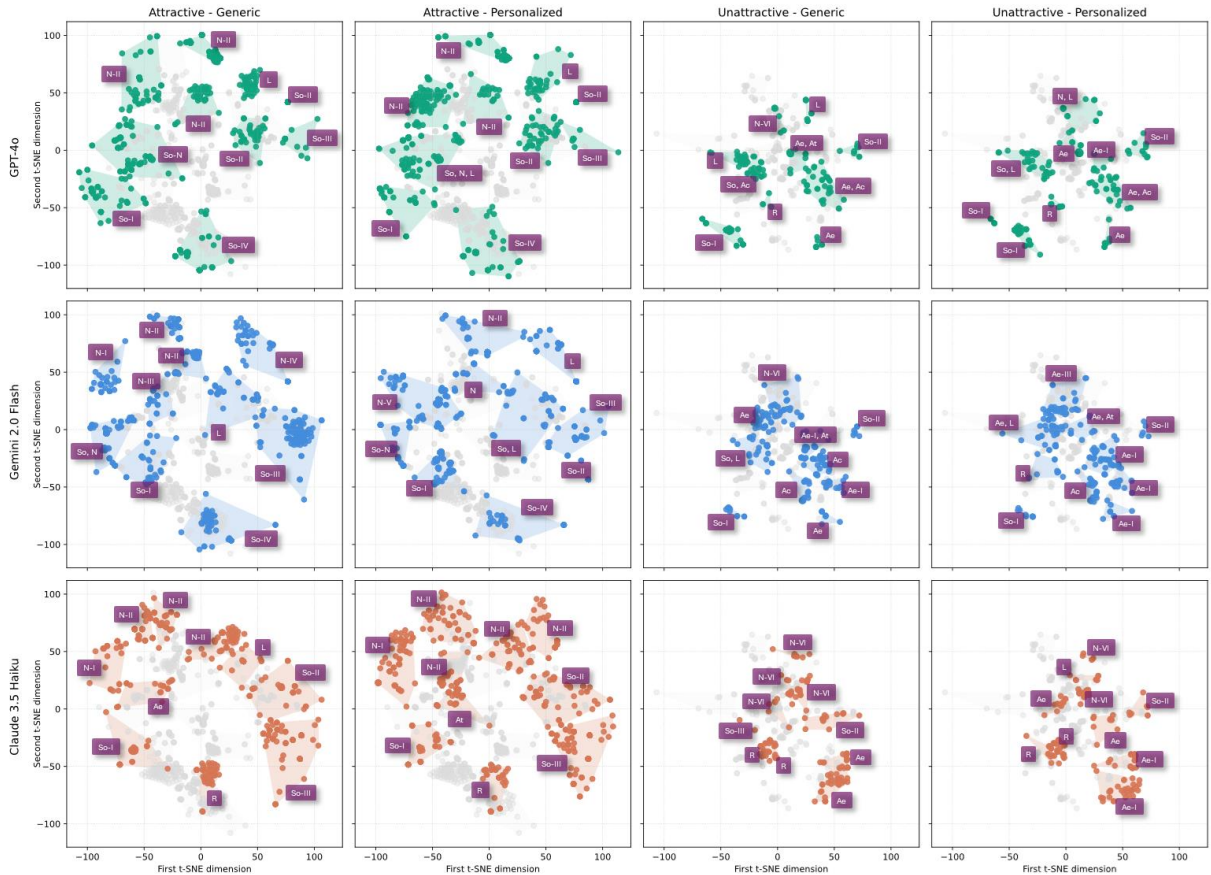

**Figure S3.** Semantic clustering of AI-generated explanations across greenspace types and prompt conditions. Each subplot shows the t-SNE projection of clustered reasons for attractive and unattractive greenspaces, evaluated by GPT-4o, Gemini 2.0 Flash, and Claude 3.5 Haiku, under generic and personalized prompt settings. Colored points represent distinct reasoning clusters, labeled using the abbreviations defined in **Table S2**, while gray points indicate the distribution of participant responses for reference.

## Supplementary Materials V – Mapping of ADE20K Classes

**Table S3.** Mapping of ADE20K classes to the aggregated categories used in both semantic and instance segmentation

| Feature | ADE20K Classes |
|---|---|
| Building | 0, 1, 25 |
| Sky | 2 |
| Road | 6, 52 |
| Greenery | 4, 9, 17, 29, 66, 72 |
| Sidewalk | 11 |
| Lighting & Signs | 43, 82, 87, 136 |
| Decorative | 22, 49, 104 |
| Furniture | 15, 19, 23, 30, 31, 33, 39, 64, 69, 137 |
| Vehicles | 20, 80, 83, 102 |
| People | 12 |

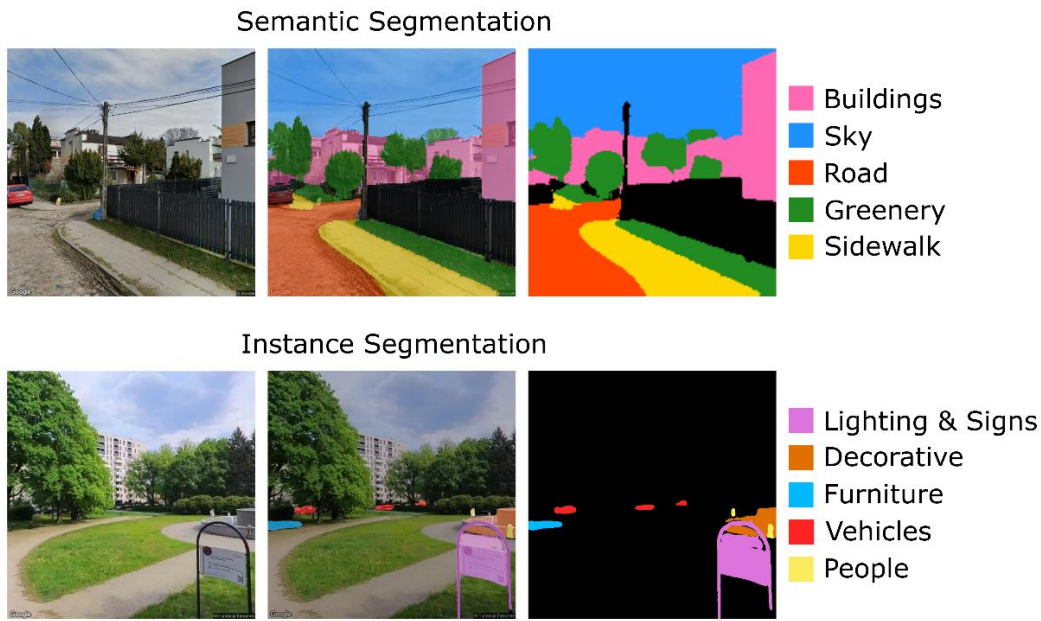

**Figure S4** - Examples of (a) semantic segmentation, and (b) instance segmentation analyses showing detection of image features.

The results of the semantic and instance segmentation analyses reveal that greenery and sky occupy the largest proportion of the image areas, reflecting the prevalence of natural and open space elements in the Street View imagery (**Table S4**). A relatively high number of vehicles are also detected, likely due to the nature of images. People, furniture, and lighting & signs are present at moderate levels, while decorative elements are much less frequently visible. The proportions of sidewalks and roads are comparatively lower than other elements.

**Table S4.** Mean and standard deviation (sd) of image features detected through semantic and instance segmentation.

| Feature | Mean (sd) |
| --- | --- |
| Road (proportion) | 0.09 (0.04) |
| Sidewalk (proportion) | 0.02 (0.02) |
| Buildings (proportion) | 0.06 (0.08) |
| Greenery (proportion) | 0.53 (0.14) |
| Sky (proportion) | 0.15 (0.07) |
| Vehicles (count) | 10.39 (13.36) |
| People (count) | 2.39 (1.66) |
| Furniture (count) | 2.66 (2.75) |
| Decorative (count) | 0.08 (0.75) |
| Lighting & signs (count) | 3.78 (4.41) |

**Supplementary Materials VI - Percentage agreement between generic and personalized prompts**

Table S5. Percentage agreement between each model and participants by greenspace type

| Model | Attractive – Formal | Attractive – Informal | Unattractive – Formal | Unattractive - Informal |
|---|---|---|---|---|
| GPT-4o | 99.5 | 97.2 | 97.6 | 97.1 |
| Gemini | 94.3 | 97.2 | 100.0 | 97.2 |
| Claude | 98.1 | 100.0 | 95.5 | 100.0 |

# Supplementary Materials VII - Distribution of reasoning categories across participants and AI models

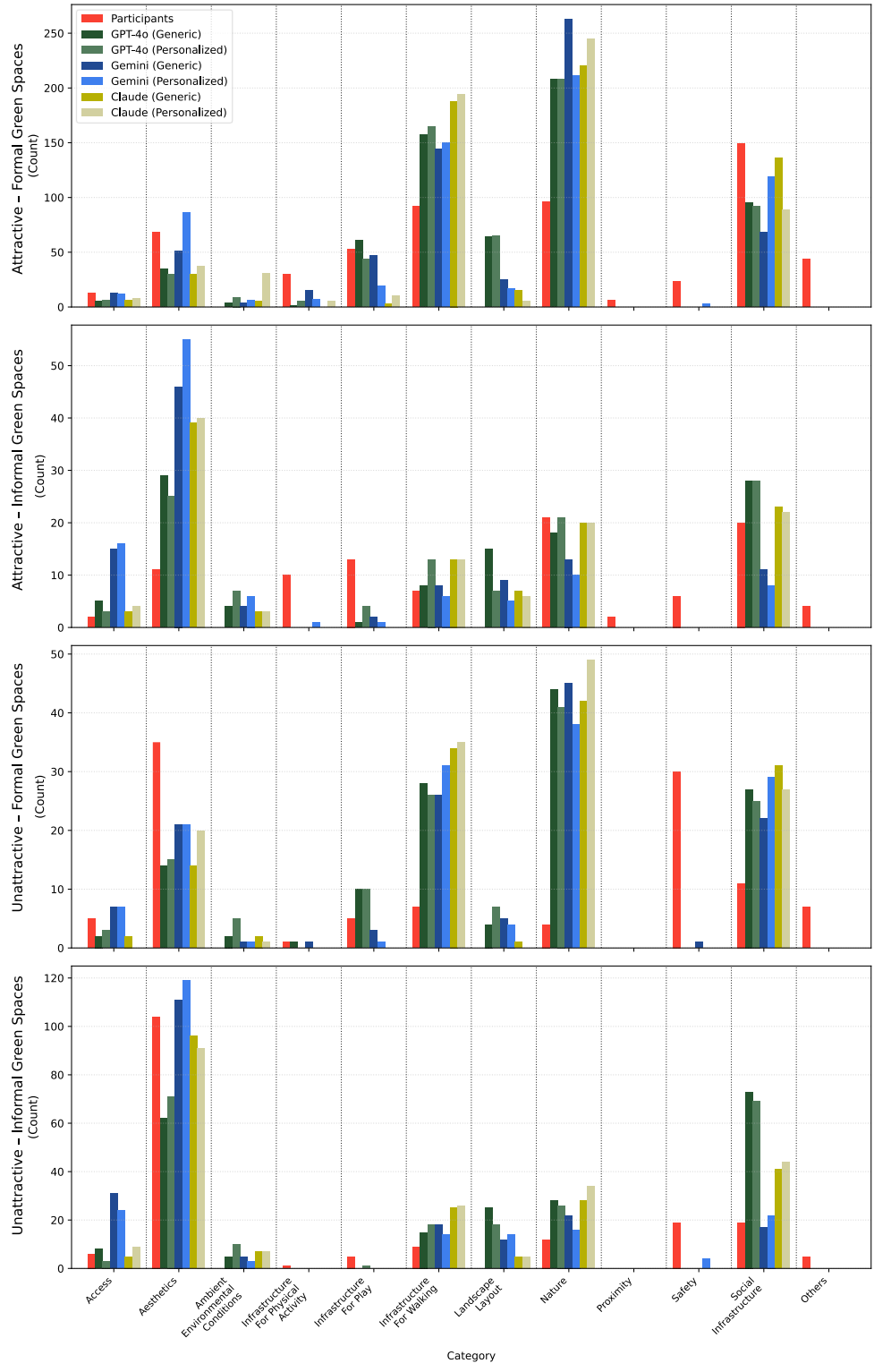

**Figure S5.** Distribution of reasoning categories across participants and AI models for attractive and unattractive greenspaces, separated by formal and informal spaces.

**Supplementary Materials IX – Agreement between AI models and participants on reasoning categories across greenspace types**

## Attractive – Formal

| | Access | Aesthetics | Ambient Environmental Conditions | Infrastructure For Physical Activity | Infrastructure For Play | Infrastructure For Walking | Landscape Layout | Nature | Proximity | Safety | Social Infrastructure | Others |
|---|---|---|---|---|---|---|---|---|---|---|---|---|
| GPT-4o (Generic) | 8.3% (1/12) | 7.0% (4/57) | | 0.0% (0/25) | 31.4% (16/51) | 73.3% (66/90) | | 79.5% (62/78) | 0.0% (0/6) | 0.0% (0/23) | 41.3% (52/126) | 0.0% (0/39) |
| GPT-4o (Personalized) | 8.3% (1/12) | 3.5% (2/57) | | 12.0% (3/25) | 25.5% (13/51) | 77.8% (70/90) | | 80.8% (63/78) | 0.0% (0/6) | 0.0% (0/23) | 42.1% (53/126) | 0.0% (0/39) |
| Gemini (Generic) | 8.3% (1/12) | 1.8% (1/57) | | 20.0% (5/25) | 17.6% (9/51) | 63.3% (57/90) | | 85.9% (67/78) | 0.0% (0/6) | 0.0% (0/23) | 29.4% (37/126) | 0.0% (0/39) |
| Gemini (Personalized) | 8.3% (1/12) | 8.8% (5/57) | | 16.0% (4/25) | 11.8% (6/51) | 66.7% (60/90) | | 79.5% (62/78) | 0.0% (0/6) | 0.0% (0/23) | 54.8% (69/126) | 0.0% (0/39) |
| Claude (Generic) | 8.3% (1/12) | 1.8% (1/57) | | 0.0% (0/25) | 5.9% (3/51) | 88.9% (80/90) | | 84.6% (66/78) | 0.0% (0/6) | 0.0% (0/23) | 65.9% (83/126) | 0.0% (0/39) |
| Claude (Personalized) | 0.0% (0/12) | 12.3% (7/57) | | 16.0% (4/25) | 3.9% (2/51) | 92.2% (83/90) | | 87.2% (68/78) | 0.0% (0/6) | 0.0% (0/23) | 46.0% (58/126) | 0.0% (0/39) |

## Attractive – Informal

| | Access | Aesthetics | Ambient Environmental Conditions | Infrastructure For Physical Activity | Infrastructure For Play | Infrastructure For Walking | Landscape Layout | Nature | Proximity | Safety | Social Infrastructure | Others |
|---|---|---|---|---|---|---|---|---|---|---|---|---|
| GPT-4o (Generic) | 0.0% (0/2) | 62.5% (5/8) | | 0.0% (0/7) | 8.3% (1/12) | 28.6% (2/7) | | 35.0% (7/20) | 0.0% (0/2) | 0.0% (0/6) | 75.0% (15/20) | 0.0% (0/3) |
| GPT-4o (Personalized) | 0.0% (0/2) | 50.0% (4/8) | | 0.0% (0/7) | 25.0% (3/12) | 57.1% (4/7) | | 50.0% (10/20) | 0.0% (0/2) | 0.0% (0/6) | 65.0% (13/20) | 0.0% (0/3) |
| Gemini (Generic) | 50.0% (1/2) | 62.5% (5/8) | | 0.0% (0/7) | 8.3% (1/12) | 28.6% (2/7) | | 10.0% (2/20) | 0.0% (0/2) | 0.0% (0/6) | 35.0% (7/20) | 0.0% (0/3) |
| Gemini (Personalized) | 0.0% (0/2) | 75.0% (6/8) | | 0.0% (0/7) | 8.3% (1/12) | 14.3% (1/7) | | 20.0% (4/20) | 0.0% (0/2) | 0.0% (0/6) | 75.0% (15/20) | 0.0% (0/3) |
| Claude (Generic) | 0.0% (0/2) | 50.0% (4/8) | | 0.0% (0/7) | 0.0% (0/12) | 57.1% (4/7) | | 35.0% (7/20) | 0.0% (0/2) | 0.0% (0/6) | 65.0% (13/20) | 0.0% (0/3) |
| Claude (Personalized) | 0.0% (0/2) | 62.5% (5/8) | | 0.0% (0/7) | 0.0% (0/12) | 42.9% (3/7) | | 35.0% (7/20) | 0.0% (0/2) | 0.0% (0/6) | 60.0% (12/20) | 0.0% (0/3) |

## Unattractive – Formal

| | Access | Aesthetics | Ambient Environmental Conditions | Infrastructure For Physical Activity | Infrastructure For Play | Infrastructure For Walking | Landscape Layout | Nature | Proximity | Safety | Social Infrastructure | Others |
|---|---|---|---|---|---|---|---|---|---|---|---|---|
| GPT-4o (Generic) | 0.0% (0/4) | 20.0% (5/25) | | 0.0% (0/1) | 40.0% (2/5) | 57.1% (4/7) | | 50.0% (2/4) | | 0.0% (0/22) | 45.5% (5/11) | 0.0% (0/7) |
| GPT-4o (Personalized) | 0.0% (0/4) | 24.0% (6/25) | | 0.0% (0/1) | 80.0% (4/5) | 28.6% (2/7) | | 50.0% (2/4) | | 0.0% (0/22) | 45.5% (5/11) | 0.0% (0/7) |
| Gemini (Generic) | 0.0% (0/4) | 20.0% (5/25) | | 0.0% (0/1) | 0.0% (0/5) | 14.3% (1/7) | | 25.0% (1/4) | | 0.0% (0/22) | 27.3% (3/11) | 0.0% (0/7) |
| Gemini (Personalized) | 25.0% (1/4) | 16.0% (4/25) | | 0.0% (0/1) | 0.0% (0/5) | 57.1% (4/7) | | 25.0% (1/4) | | 0.0% (0/22) | 45.5% (5/11) | 0.0% (0/7) |
| Claude (Generic) | 25.0% (1/4) | 28.0% (7/25) | | 0.0% (0/1) | 0.0% (0/5) | 71.4% (5/7) | | 50.0% (2/4) | | 0.0% (0/22) | 45.5% (5/11) | 0.0% (0/7) |
| Claude (Personalized) | 0.0% (0/4) | 28.0% (7/25) | | 0.0% (0/1) | 0.0% (0/5) | 71.4% (5/7) | | 50.0% (2/4) | | 0.0% (0/22) | 45.5% (5/11) | 0.0% (0/7) |

## Unattractive – Informal

| | Access | Aesthetics | Ambient Environmental Conditions | Infrastructure For Physical Activity | Infrastructure For Play | Infrastructure For Walking | Landscape Layout | Nature | Proximity | Safety | Social Infrastructure | Others |
|---|---|---|---|---|---|---|---|---|---|---|---|---|
| GPT-4o (Generic) | 0.0% (0/5) | 71.7% (43/60) | | 0.0% (0/1) | 0.0% (0/5) | 22.2% (2/9) | | 45.5% (5/11) | | 0.0% (0/18) | 100.0% (17/17) | 0.0% (0/4) |
| GPT-4o (Personalized) | 0.0% (0/5) | 78.3% (47/60) | | 0.0% (0/1) | 0.0% (0/5) | 33.3% (3/9) | | 27.3% (3/11) | | 0.0% (0/18) | 94.1% (16/17) | 0.0% (0/4) |
| Gemini (Generic) | 80.0% (4/5) | 83.3% (50/60) | | 0.0% (0/1) | 0.0% (0/5) | 33.3% (3/9) | | 9.1% (1/11) | | 0.0% (0/18) | 23.5% (4/17) | 0.0% (0/4) |
| Gemini (Personalized) | 60.0% (3/5) | 86.7% (52/60) | | 0.0% (0/1) | 0.0% (0/5) | 44.4% (4/9) | | 9.1% (1/11) | | 0.0% (0/18) | 41.2% (7/17) | 0.0% (0/4) |
| Claude (Generic) | 20.0% (1/5) | 75.0% (45/60) | | 0.0% (0/1) | 0.0% (0/5) | 66.7% (6/9) | | 9.1% (1/11) | | 0.0% (0/18) | 64.7% (11/17) | 0.0% (0/4) |
| Claude (Personalized) | 20.0% (1/5) | 78.3% (47/60) | | 0.0% (0/1) | 0.0% (0/5) | 77.8% (7/9) | | 27.3% (3/11) | | 0.0% (0/18) | 82.4% (14/17) | 0.0% (0/4) |

**Figure S6**. Agreement between AI models and participants on reasoning categories across greenspace types. Each cell shows the percentage of cases where an AI model selected the same reasoning category as participants for a given greenspace, reported separately for attractive and unattractive, formal and informal spaces. In brackets, the absolute counts indicate how many times the model matched the participant's reasoning over the total number of times participants mentioned that category. Darker shades indicate higher levels of agreement. Columns are shown in gray where participants did not mention the category for that environment.

**Supplementary Materials X– Co-occurrence of reasoning categories in AI-generated explanations**

Across all models, the categories Nature, Infrastructure for Walking, Aesthetics, and Social Infrastructure co-occur most frequently with other reasoning categories. This reflects their overall dominance in the explanations. Most categories, regardless of their specific theme, frequently appear alongside at least one of these four categories, suggesting that models tend to describe greenspace attractiveness or unattractiveness using a combination of these common factors rather than isolated reasons.

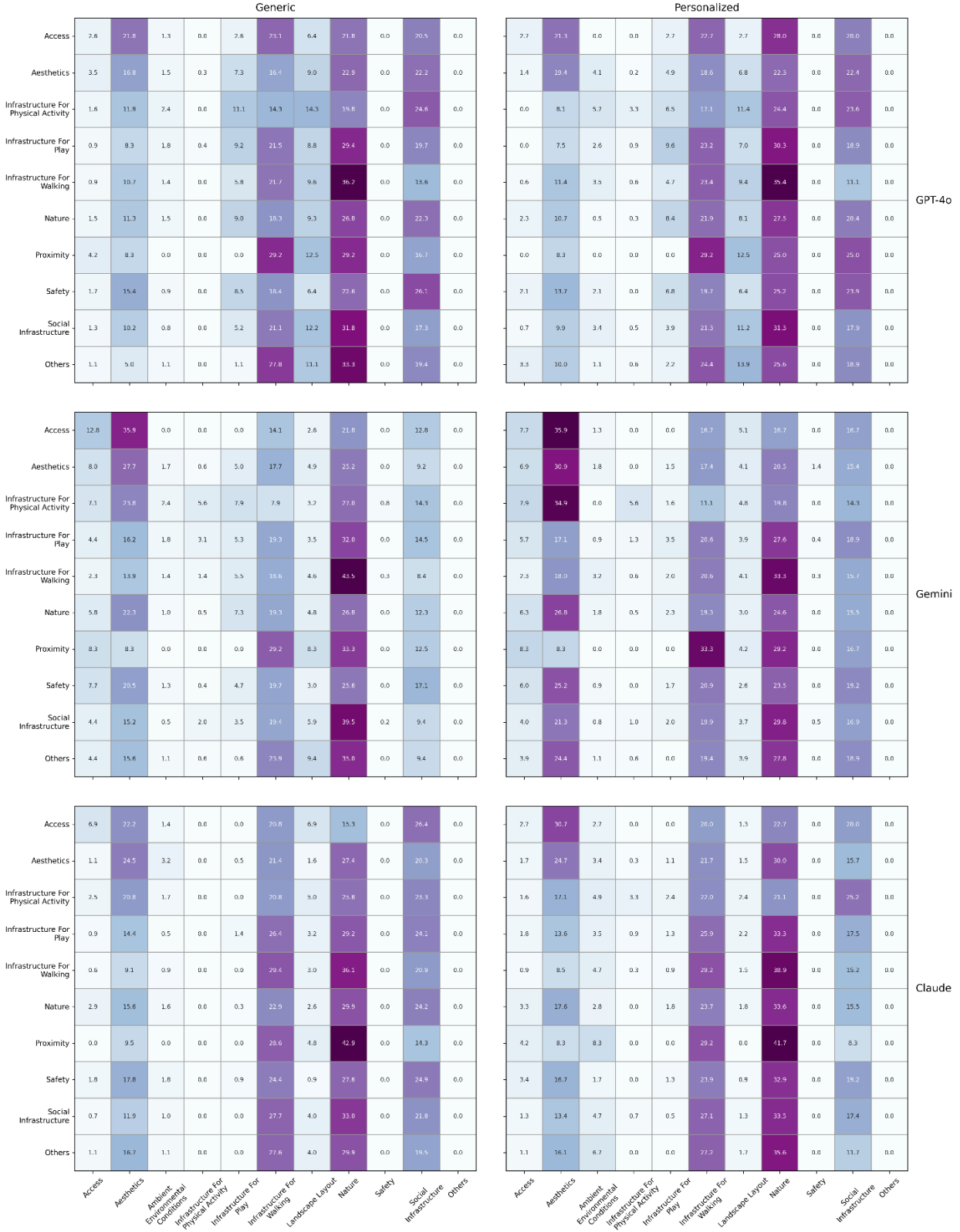

**Figure S7**. Co-occurrence of reasoning categories in AI-generated explanations. Each row represents a reasoning category, and the cells indicate how frequently other categories co-occur with it, normalized by the total occurrences of that row's category. Darker shades indicate higher co-occurrence.

**Supplementary Materials XI – Distribution of image features across reasoning categories**

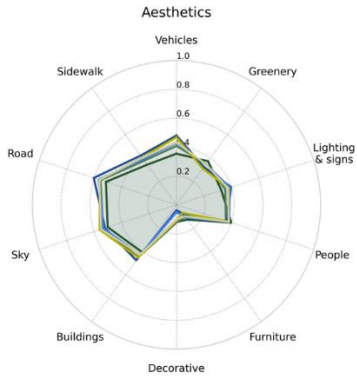
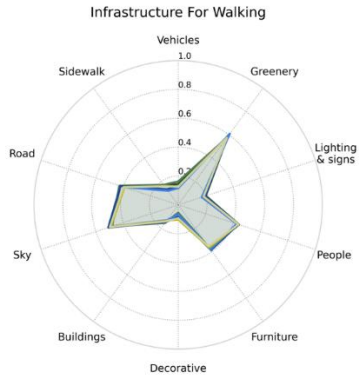
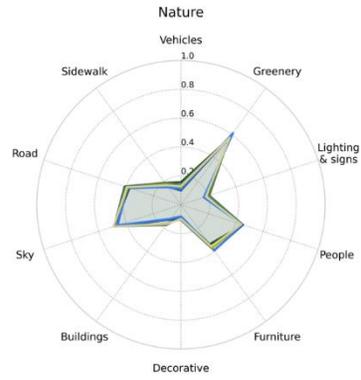
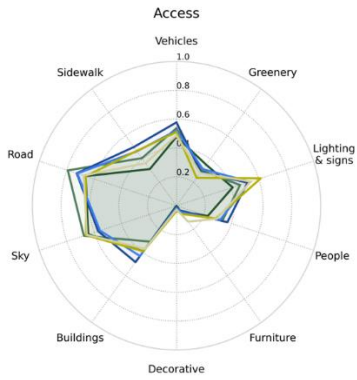
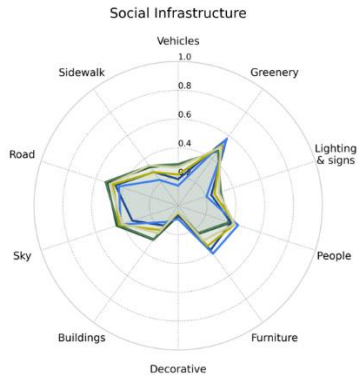
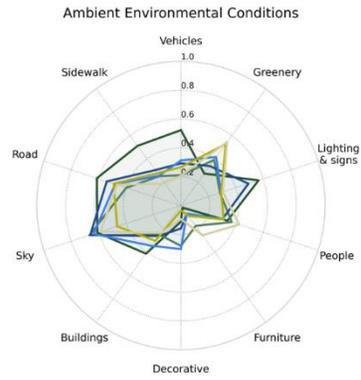
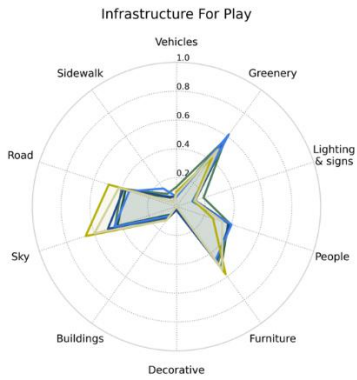
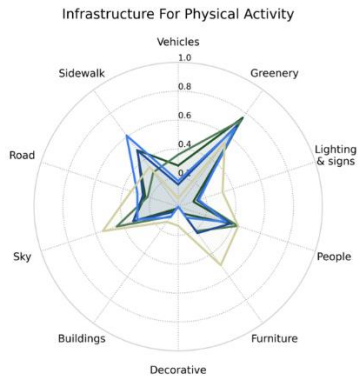
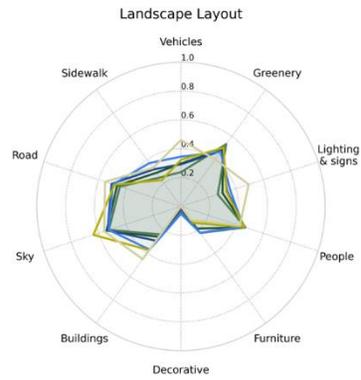
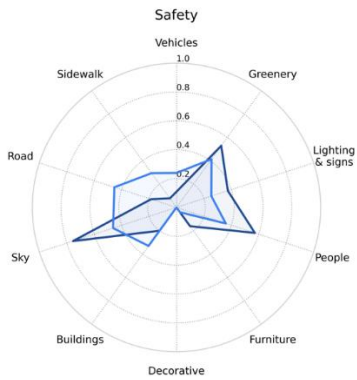
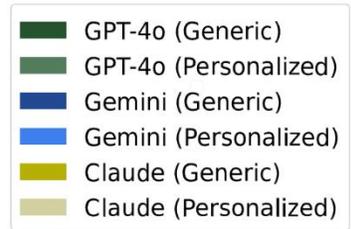

**Figure S8.** Distribution of image features across reasoning categories. Each radar chart shows the normalized average proportion of visual features detected in the images linked to each reasoning category, as interpreted by the AI models.